\documentstyle[multicol,aps,epsf]{revtex}
\begin{document}
\draft
\title{Incommensurability in the Frustrated Two-Dimensional $XY$ Model}
\author{Colin Denniston$^{1,2,}$\cite{present} and Chao Tang$^2$}
\address{$^{1}$Department of Physics, Princeton University, 
Princeton, New Jersey 08544}
\address{$^{2}$NEC Research Institute, 4 Independence Way, Princeton,
New Jersey 08540}

\date{\today}
\maketitle

\begin{abstract}
To examine the properties of the frustrated $XY$ model at an 
incommensurate field we have examined a sequence of magnetic filling factors 
$f$ which approach the irrational value of one minus the golden mean.  
At all $f$ studied, the system undergoes a finite-temperature weak first-order
transition involving the freezing out of Ising-like domain walls.
As one approaches incommensurability, the low temperature phase of the system 
changes from the staircase states found by Halsey \cite{Halsey} to a striped 
phase consisting of a superlattice of parallel shift (Pott's-like) domain 
walls.
\end{abstract}

\begin{multicols}{2}
The frustrated $XY$ model provides a convenient framework to study a
variety of fascinating phenomena displayed by numerous physical systems. 
One experimental realization of this model is in two-dimensional
arrays of Josephson junctions and superconducting wire networks
\cite{jja,us,us2,Sean}.  A perpendicular magnetic field induces a finite
density of circulating supercurrents, or vortices, within the array.  The
interplay of two length scales -- the mean separation of vortices and the
period of the underlying physical array -- gives rise to a wide variety
of interesting physical phenomena.  Many of these effects show up as
variations in the properties of the finite-temperature superconducting
phase transitions at different fields.  Recent and ongoing 
experiments have been able to make measurements of the critical 
exponents in superconducting arrays \cite{Sean}, opening the opportunity 
to do careful comparison of theory and experiment.  In this Letter we examine 
the critical properties of the 2D $XY$ model in the densely frustrated regime 
($f \gg 0$) and for a sequence of fields which approach incommensurability.

The Hamiltonian of the frustrated $XY$ model is
\begin{equation}
{\cal H} = - J \sum_{\langle ij \rangle} \cos(\theta_i-\theta_j-A_{ij}),
\label{ham}
\end{equation}
where $\theta_j$ is the phase on site $j$ of a square $L \times L$ lattice and
$A_{ij}=(2\pi/\phi_0)\int^j_i {\bf A} \cdot d{\bf l}$ is the integral of the
vector potential from site $i$ to site $j$ with $\phi_0$ being the flux 
quantum. The directed sum of the $A_{ij}$ around an elementary plaquette 
$\sum A_{ij}=2\pi f$ where $f$, measured in the units of $\phi_0$, is the
magnetic flux penetrating each plaquette due to the uniformly applied
field.

It is, in general, quite hard to study a system near incommensurability.  For
numerical work this is due to the large system sizes required; for $f=p/q$ a
system of {\it at least} $q \times q$ is required, and by definition 
$q \rightarrow \infty$ for incommensurate $f$.  A lack of knowledge about 
the ground state of the system and the low energy excitations makes a 
phenomenological description of the phase transitions quite 
difficult.  It has been speculated that as one approaches 
irrational $f$, some ``soft-mode'', of unknown form and origin, will enter 
the system resulting in the transition temperature approaching zero 
\cite{Halsey2,TeitJay,Granato}.  
Some works, based on Monte-Carlo simulations of systems near 
incommensurability, have suggested the system freezes at a finite temperature
to a glass state \cite{Halsey2,Granato}.  
Our work suggests that this glassy behavior is due to the periodic 
boundary conditions used in these simulations.  We find that, even near
irrational $f$, the system has a finite temperature transition to an
ordered state.

To study the system near incommensurability we examine a sequence of states, 
$f=3/8$, $5/13$, $8/21$, $13/34$, $21/55$, $34/89\cdots$ which approaches the 
quadratic irrational value of $1-\Omega$ $=$ $(3-\sqrt{5})/2$ 
($\Omega$ $=$ $0.618\cdots$ is the golden mean).  We examine ground state 
properties and low energy excitations using a numerical constrained 
optimization to 
minimize the energy.  We correlate these states to those found with
Monte Carlo (MC)
 simulations of systems with {\it soft} boundary conditions (soft boundary 
conditions are necessary to relax the constraint of fixed periodicity).  We find
 that these systems can have low temperature striped phases similar to those 
found in com\-mensurate-incom\-mensurate transitions.  For all $f$
studied we find a finite temperature 
phase transition which does not appear to approach zero. 

\begin{figure}
\narrowtext
\centerline{\epsfxsize=3.2in
\epsffile{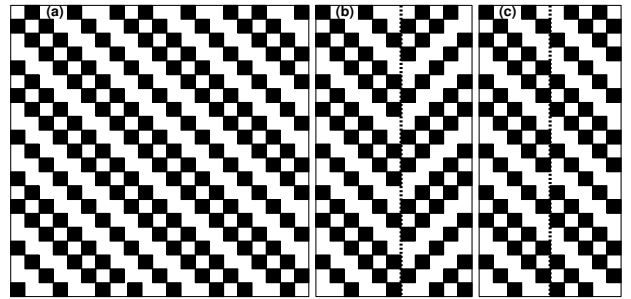}}
\vskip 0.1true cm
\caption{Fluxoid pattern for $f={8\over 21}$ for (a) a unit cell of the
staircase state, (b) A herringbone wall, and (c) and shift-by-eight
wall. A vortex is shown as a dark square. In (c), the pattern on the right
is shifted down by eight from where it would be if it had just continued 
the pattern on the left.}
\label{f821walls}
\end{figure}

The ground states of the Hamiltonian (\ref{ham}) will be among the solutions to
the supercurrent conservation equations 
$\partial {\cal H}/\partial \theta_i=0$:
\begin{equation}
\sum_{j'} \sin(\theta_{j'}-\theta_i-A_{ij'})=0
\label{curr_cons}
\end{equation}
where $j'$ are the nearest neighbors to $i$.  One set of solutions to these
equations was found by Halsey \cite{Halsey} in which the square network is
partitioned into diagonal staircases with a constant current flowing along
each staircase. 
The resulting fluxoid patterns consist of diagonal lines of vortices. A unit 
cell of the staircase fluxoid pattern for $f=8/21$ is shown in 
Fig~\ref{f821walls}(a).  For $f=p/q$, 
the staircase pattern can sit on $q$ sub-lattices and, in addition, there are 
$q$ states with the lines of vortices going along the opposite diagonal, making
 a total of $2 q$ degenerate states ($f=p/q \ne 1/2$).  While these staircase 
states are not, as will be shown later, the ground states for all $f$,
they are a useful set of states 
in that the striped phases we find near $f=\Omega$ can be defined in terms of 
domains of staircase states separated by parallel domain walls. 

\begin{figure}
\narrowtext
\centerline{\epsfxsize=3.25in
\epsffile{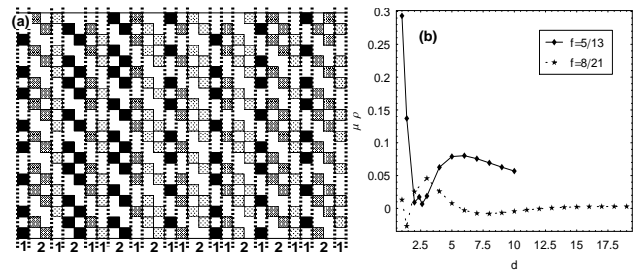}}
\vskip 0.1true cm
\caption{(a) A section of the striped phase for $f=8/21$ corresponding to the
minimum in the energy in (b).  The sequence of wall spacings repeats 
periodically.  The shading is only a guide for the eye. (b) Energy per unit 
area of a superlattice of shift-by-eight walls as a function of their average 
separation for $f$ $=$ ${5\over 13}$ and ${8\over 21}$.}
\label{stripe}
\end{figure}

Figure~\ref{f821walls}(b) and (c) show some of the low energy domain walls for 
a typical case, $f=8/21$.  Domain walls between the $2q$ degenerate
staircase states can be classified into two types. 
{\it Shift} walls involve a shift of the vortex pattern across the wall (such 
as in Fig~\ref{stripe}(c) where the pattern on the right is shifted down by 8 
lattice spacings with respect to the pattern on the left) but the lines of 
vortices are still going along the same diagonal.  {\it Herringbone} walls are 
walls between states with the vortex lines going along opposite diagonals.  

To calculate energies of different vortex patterns, we solved
equations (\ref{curr_cons}) numerically, using a quasi-Newton method, on 
lattices with up to $2.3\times10^5$ sites and constraints fixing the fluxoid 
occupation of each plaquette.  For $f=3/8$ the lowest energy
wall is a herringbone wall, but there is a shift wall with only slightly
higher energy.  For the higher order rationals ($f=5/13$ to $34/89$) there 
is at least one shift wall with lower energy than any single herringbone wall. 
In addition, a striped phase such as the one shown in Fig.~\ref{stripe}(a), 
consisting of a super-lattice of parallel shift walls is lower in energy than 
the plain staircase state for $f=8/21$, $13/34$, $21/55$ and $34/89$.
 
The energy of a superlattice of parallel shift walls is shown in
 Fig~\ref{stripe}(b) for $f=5/13$ and $8/21$.  The presence of  
 a short-range energetic repulsion makes wall crossings energetically 
unfavorable.  Also, the interaction is essentially flat at large distances.  
There is, however, a minimum in the 
interaction potential at a finite separation of the walls.  This minimum 
arises due the directionality of the wall which causes the distortion of the 
phase to be asymmetric on each side of the wall.  As a result, when the 
distortion of the phases from two walls overlap, there can be some cancellation.
  If however, the walls get too close, the distortions of the phase field from 
the two walls start to match, causing a rapid increase in energy.

\begin{figure}
\narrowtext
\centerline{\epsfxsize=3.25in
\epsffile{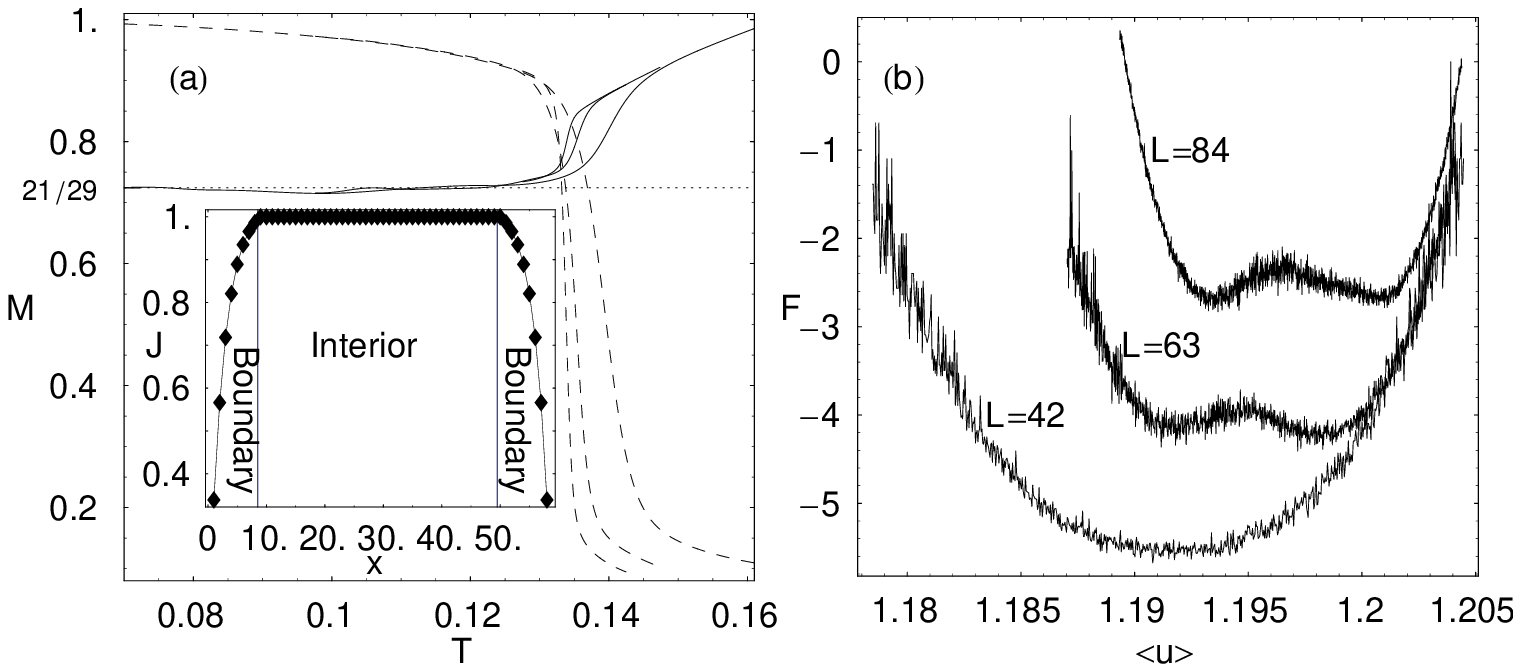}}
\vskip 0.1true cm
\caption{(a) Diagonal order (dashed) and shift wall density (solid) versus 
$k_B T/J$ for $f=8/21$ and $L=42, 63,$ and $84$. Dotted line indicates
a shift wall density of 21/29.  Inset: Couplings $J$ in a 
cross-section of the system for $L=42$.  Data from the boundary layers is 
discarded.  (b) Free energy barrier between ordered and disordered state 
for $f=8/21$.}
\label{f821_mc}
\end{figure}

While energy calculations show that the striped phase is lower in energy than
the plain staircase state for $f=8/21$, $13/34$, $21/55$ and $34/89$, this does
not necessarily mean that there is not some other state with even lower energy.
To test this, we undertook extensive Monte Carlo 
simulations.  Boundary effects can propagate quite far into the system
as the boundary can easily induce a high energy domain wall into the system. 
To alleviate this strain, the system will break this single high energy 
interface into numerous lower energy walls resulting in a complicated structure
of walls which can extend well into the system.
These trapped domain structures will have a 
significant effect on finite temperature states and transitions found in 
smaller samples. Free boundary conditions can also induce domain walls, as a 
free boundary can act like a mirror plane in the system.
The problems of these boundary conditions can be alleviated by 
installing a boundary layer at the edge of the system.  In the boundary layer, 
the coupling $J$ in (\ref{ham}) 
goes from one on the interior side to zero on the exterior: 
$J(x)=A(1-e^{-x/\lambda})$ where $A=1/(1-e^{-w\lambda})$ and $w$ is the number
of rows of lattice sites in the boundary layer and $\lambda<w$ is adjusted so 
that connection to the interior (where $J=1$) is reasonably smooth 
(see inset on Fig.~\ref{f821_mc}(a)).  We found that $w=8$ and $\lambda=2.5$ 
gave reasonable results.  In practice, it is only necessary to use boundary 
layers in one direction and periodic boundary conditions can be kept in the 
other.  This does lead to a preferred direction in 
the striped phase (stripes parallel to the boundary layers) but does not seem 
to have a qualitative effect on the system behavior (it does however affect 
finite size effects).  Measurements of the energy, order parameter, etc. were
made only on the interior, where the coupling $J=1$.

For the discrete degrees of freedom we kept track of an orientational order
parameter $M_d$, measuring whether the vortices are preferentially
arranged along one diagonal, and in the striped phase an
order parameter $\rho$ measuring the density of shift walls.  The
MC simulations used a heat bath algorithm with system sizes 
$32$ $\leq$ $L$ $\leq$ $96$.  We computed about $10^7$ MC
steps (complete lattice updates), and data from different temperatures
was combined and analyzed using histogram techniques \cite{FerrenburgSwend}.  

At the lowest temperatures of the simulations, $k_BT/J=0.03$, we find the 
system goes into the states expected from the energy calculations:
$f=3/8$ and $5/13$ are in the plain staircase states and $f=8/21, 13/34,$ and 
$21/55$ are in a striped phase.  In the following discussion, we start by 
examining $f=8/21, 13/34,$ and $21/55$ which undergo a first order phase 
transition at about $k_BT_c$ $=$ $0.13 J$ from the striped phase to the 
diagonally disordered phase.  We then turn to the $f$ $=$ $5/13$ case which has
 a transition from the plain staircase state to the striped phase at 
$k_BT_c$ $=$ $0.04 J$ and then has a transition at $k_BT_c$ $=$ $0.13 J$ to a
diagonally disordered phase.  For the largest system sizes, 
$f$ $=$ $3/8$ appears to undergo a single transition from the plain staircase 
state to the diagonally disordered phase at about $k_BT_c$ $=$ $0.123 J$.

Figure~\ref{f821_mc}(a) shows the diagonal order
and shift wall density as a function of temperature for $f=8/21$.  In the high 
temperature phase, domain walls of all types, shift and herringbone, are 
present and the vortex lattice is disordered.   At the critical 
temperature $T_c$, the system orders by freezing out herringbone walls, 
leaving a diagonally ordered striped phase.  
This striped phase has a density $\rho=21/29$ of 
shift walls almost independent of temperature (see Fig.~\ref{f821_mc}(a)). 
($\rho=21/29$ corresponds to an average spacing of $1 {8\over 21}$.)   
The non-integer spacing comes from a mixture of wall spacings $d_i$ of 1 and 2,
  arranged in a Fibonacci sequence cut off at 21,
\begin{eqnarray}
d_0&=&1,\nonumber\\
(d_1,d_2)&=&(2,1),\nonumber\\ 
(d_3,d_4,d_5)&=&(d_0,(d_1,d_2)),\nonumber\\ 
(d_6,d_7,d_8,d_9,d_{10})&=&((d_1,d_2),(d_3,d_4,d_5)),\nonumber\\
\cdots d_{21}.
\label{spacings}
\end{eqnarray}
Note that it is the wall spacings that repeat periodically every 21 walls, but
the actual vortex lattice period repeats every $29\times 21=609$ lattice 
constants (one period of the wall spacings takes 29 lattice constants for 
$\rho=21/29$).  Thus, the typical period of the 
ground state vortex lattice can be of order $q^2$ rather that $q$ for $f=p/q$.
The spacing observed in the Monte Carlo simulations corresponds to 
the system sitting at the minimum of the energy in Fig.~\ref{stripe}(b).
We should note that this is quite different from the normal case studied in 
commensurate-incommensurate transitions \cite{Pokrovsky} where there is no 
minimum in the interaction potential to pin the walls.

The transition from the striped phase to the diagonally disordered phase 
appears to be first order.  This is indicated by the 
presence of a free energy barrier at the 
transition between the ordered and disordered states which diverges as the 
system size increases \cite{LeeKosterlitz}.  The free energy as a function of 
energy is obtained using ${\cal F}_L(E)=-\ln P_L(E)$ where $P_L(E)$ is the 
probability distribution for the energy generated by Monte Carlo simulation of 
a $L\times L$ system. Figure~\ref{f821_mc}(b) shows the growth in this barrier 
as the system size increases from $L=42$ to $84$ giving evidence for a first 
order transition.  The barrier is, however, growing quite slowly so the 
transition is only weakly first order and the system sizes available are not 
large enough to apply finite size scaling to confirm the nature of the 
transition.  One can do an approximate extrapolation of the measured $T_c(L)$ 
to obtain $T_c$ $=$ $0.1325\pm 0.0007$ for $f=8/21$.  

The $f=13/34$ and $21/55$ cases also undergo what appears to be a first order 
phase transition at around the same $T_c=0.13$ from the 
diagonally disordered high temperature phase to the striped phase.  The striped
 phases for these $f$ appears to be slightly more complicated.  Like the 
$f=8/21$ case, the stripes appear to be mainly composed of shift-by-eight 
walls with a similar Fibonacci sequence of spacings (with spacings of 2 
and 3).  However, these higher order rationals also have other walls which have
negative energy with respect to the staircase state.  These additional walls
also seem to be present in a much lower density, inter-spaced between the 
shift-by-eight walls in some quasi-periodic pattern.  If one includes these 
walls, the wall density is similar to $f$ $=$ $8/21$ and the vortex lattices 
look very similar to $f$ $=$ $8/21$, which is probably why they have such 
similar $T_c$.  
  
For $f=5/13$ the shift walls are the lowest energy walls, but the striped phase
costs energy (see Fig.~\ref{stripe}(b)).  The striped phase
can exist at finite temperature however, due to entropic reasons which we shall
discuss below. In the Monte Carlo simulations we see a very similar transition 
for $f$ $=$ $5/13$ (similar $T_c$ and weak first order) to the one seen for 
$f$ $=$ $8/21$ from the 
diagonally disordered high temperature phase to the striped phase.  The 
wall density in the striped phase is fixed at about $\rho$ $=$ $13/31$, which
can be constructed from a Fibonacci sequence of wall spacings consisting of
spacings of $2$ and $3$ in a manner similar to that used in 
Eq.(\ref{spacings}).  For $L$ $=$ $39$, and at about $T$ $\approx$ $0.05$ the 
wall spacings of $2$ and $3$ appear to switch to give a slightly lower energy 
state at $\rho$ $=$ $13/34$.  It is unclear however if this would be the case 
for larger systems and would require further study.  These two wall densities, 
$\rho$ $=$ $13/34$ and $13/31$ correspond to the two dips within the larger
minima seen in the interaction energy shown in Figure~\ref{stripe}(b).  At a 
lower temperature $T_p$ $\approx$ $0.045$, the system undergoes another first 
order transition from the striped phase to the plain staircase state.  

The transition from the plain staircase state to the striped phase is 
similar to the commensurate-incommensurate transitions studied in the context
of adsorbed films \cite{Pokrovsky}, which is a second order phase 
transition.   In studies of these transitions, one considers the free energy of
 a single line per unit length $\epsilon_s$.  This can be 
estimated using a simple solid-on-solid (SOS) model of the shift line. The 
energy of the line, extending from one side of the system to the other is
\begin{equation}
{\cal H}_s \{z\}=\sigma_\parallel L+\sigma_\perp \sum_k |z_k-z_{k-1}|.
\label{SOSham}
\end{equation}
where $\sigma_\parallel$ ($\sigma_\perp$) is the energy per unit length in the
direction parallel (perpendicular) to the wall.  The heights $z_k$, 
take on integer values.  The partition function, 
can be evaluated to give the interfacial free energy per 
column \cite{Forgacsetal}
$\epsilon_s=T \ln [ e^{\sigma_\parallel/T}\tanh(\sigma_\perp/(2T))].$
A phase transition occurs when $\epsilon_s$ becomes negative.  If this were
the case here, one would see a continuous rise in the shift-wall density.
What makes this case different is the presence of the minimum in the wall 
interaction potential (Fig.~\ref{stripe}(b)).  

If placed in a system with other shift walls, the walls will experience an
entropic repulsion since a wall can only occupy the region of space between 
it's neighbors.  To see whether or not this entropic repulsion is relevant,
we estimate if two walls remain bound together at the minima of the 
interaction potential.  This is done using a SOS model for two walls with 
a binding energy equal to the depth of the minima in the interaction:
\begin{eqnarray}
{\cal H}_d \{\Delta,z\}= 
	\sum_k \{(2b\sigma+u_\parallel\delta_{z_k,0})
	+b\sigma |z_k-z_{k-1}|\nonumber\\
	 +(2b\sigma+u_\perp \delta_{z_k,0}) \Delta_k\}.
\label{HdoubleSOS}
\end{eqnarray}
where $z_k$ is the separation of the walls ($z_k \geq 0$), and
$\Delta_k$ is the number of vertical steps the two walls take in the same
direction in the k'th column ($-\infty < \Delta_k <\infty$).  $u_\parallel$
and $u_\perp$ are the binding energies parallel and perpendicular
to the wall.  Summing over 
$\Delta_k$ leaves the partition function in the form of a transfer 
matrix: ${\cal Z}$ $=$ $\sum_{\{z_k\}} \prod_k T_{z_k}^{z_{k-1}}$.
A ground state 
eigenvector $\psi_\mu(z)$ $=$ $e^{-\mu z}$, where $1/\mu$ is the localization 
length, or typical distance separating the lines, characterizes the bound state
 of the two lines.  $\mu=0$ defines the unbinding transition at $T_b$.
Doing this, 
one finds an unbinding temperature of $k_B T_b/J=0.51$.  Below $T_b$, the 
entropic repulsion is insufficient to push the system out of the minimum.  
Above $T_b$, the striped ``solid'' phase will melt into a phase where the wall 
density changes continuously with temperature.  Here, however, this is 
preempted by the entrance of the diagonally disordered phase at $T_c=0.13 J$.
    
In order for the striped phase to be stable for $f=5/13$, where it costs
energy, there must be sufficient entropy from the lines wandering within the 
region between it's neighbors.
The energy per line at finite temperature can be 
estimated using (\ref{SOSham}) with $z_k$ restricted 
to $0,\pm 1$ as the minimum in the interaction energy is at a spacing of 
about 2.  The free energy per line per column is then
$$
\epsilon_b=T \ln[e^{\sigma_\parallel/T}/(1+e^{-2 \sigma_\perp/T}(1+\sqrt{1+8 e^{2 \sigma_\perp/T}})/2)].
$$
At the point where $\epsilon_b$ crosses zero the striped phase coexists with 
the plain staircase state and a first order phase transition occurs.  Taking 
 $\sigma_\parallel=0.041 J$ for the shift-by-eight wall at the minimum of 
the interaction energy and $\sigma_\perp=0.04 J$ 
from an average of measurements of the energy of several kinks of differing 
lengths, one finds that $\epsilon_b$ crosses zero at $T_b=0.04 J$ in 
reasonable agreement with the value observed in the Monte Carlo simulations.

In experiments \cite{Sean}, a
finite temperature second order phase transition is seen at $f=\Omega$.  
That the transition occurs at finite temperature is in agreement with our
results but the continuous transition appears to disagree
with the very weak first order phase transition found here.  However, bond 
disorder which is always present experimentally, has been shown to
wipe out any coexistence region of two phases in two dimensions making
all transitions continuous \cite{Imry-Ma-Aizenman-Wehr-Hui-Berker,us2}.  

In conclusion, we find that all of the systems studied undergo a finite 
temperature first order transition from an ordered state to a diagonally
disordered state.  The transition temperature is nearly constant and shows no
signs of approaching zero as one goes to more incommensurate $f$.  As one 
approaches incommensurate $f$,  the low
temperature state changes from the plain staircase state found by Halsey to
the striped phase.

\end{multicols}

\end{document}